\documentclass{pasj00}
\draft

\begin{document}
\SetRunningHead{Yutaka Fujita}{Does a Galaxy Fly?}
\Received{2006/04/24}
\Accepted{2006/08/14}

\title{Does a Galaxy Fly?}

\author{Yutaka \textsc{Fujita}}
\affil{Department of Earth and Space Science, Graduate School of
Science, \\
Osaka University, Toyonaka, Osaka 560-0043}
\email{fujita@vega.ess.sci.osaka-u.ac.jp}


\maketitle

\begin{abstract}
Disk galaxies in a cluster of galaxies are moving in hot gas filling the
cluster. Generally, they are moving at transonic or supersonic
velocities. If ram-pressure stripping is insufficient to destroy the gas
disk, the galaxies should be affected by the wind of the surrounding hot
gas similar to an airfoil. In this paper, I consider whether the
aerodynamic interaction can be strong enough to force a disk galaxy to
deviate from the orbit that it would have been in. I find that while the
lift force is not effective, the drag force could affect face-on disk
galaxies in poor clusters on long timescales.
\end{abstract}

\section{Introduction}

The hot gas in clusters is often called the intracluster medium
(ICM). The typical temperature and density of the ICM are $T_{\rm
ICM}\sim 2$--10~keV and $\rho_{\rm ICM}\sim 10^{-27}\rm\: g\: cm^{-3}$,
respectively \citep{sar86}. While most galaxies are moving at transonic
velocities ($V\sim 1000\rm\: km\: s^{-1}$) \citep{sar86}, the galaxies
that fall into a cluster from the outside have a Mach number of ${\cal
M}\sim 2$--3 \citep{ghi98}. So far, the interaction between disk
galaxies and the surrounding ICM has been considered mostly in terms of
`ram-pressure stripping' \citep{gun72,fuj99,fuj99b,got03}. The disk of a
disk galaxy is composed of stars and gas filling the disk. The latter is
called interstellar medium (ISM), and is cold and dense ($T_{\rm
ISM}\lesssim 10^4$~K and $\rho_{\rm ISM}\gtrsim 10^{-24}\rm\: g\:
cm^{-3}$) compared with the ICM. The ISM prevents the penetration of the
galaxy by the ICM. If the ram-pressure force of the ICM ($\sim \rho_{\rm
ICM}V^2$ per unit area) becomes larger than the gravitational restoring
force of the galaxy, the ISM is stripped away from the galaxy
\citep{gun72}. Once the ISM is stripped, the ICM flows through the
stellar disk of the galaxy without much resistance. Many numerical
simulations have been performed to study this ram-pressure stripping
\citep{aba99,mor00,qui00,sch01,vol01,mar03,roe05a,roe05b}.  Another
possible interaction between the ICM and ISM may be aerodynamic effects
on a galaxy by the flow around the galaxy. In this paper, I discuss a
few cases where the aerodynamic forces could operate on the galaxy
motion, assuming that a disk galaxy is a thin solid disk for
simplicity. Since the aerodynamic effects are available only when the
disk ISM is not completely removed, I discuss only the cases where
numerical simulations showed that the ram-pressure does not remove the
ISM completely. Ram-pressure stripping is ineffective when the attack
angle of a galaxy is small, and/or when the galaxy velocity is small. I
consider the former in section~\ref{sec:lift} and the latter in
section~\ref{sec:drag}.

\section{Lift Force}
\label{sec:lift}

\begin{figure}
  \begin{center}
    \FigureFile(80mm,80mm){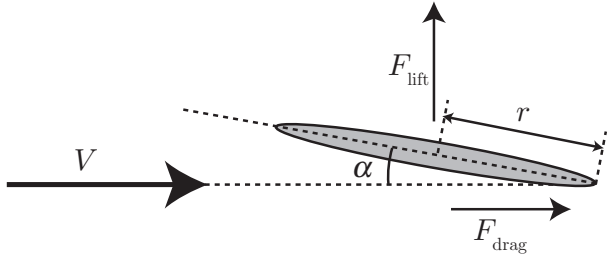}
  \end{center}
  \caption{Cross section of the galaxy. The galaxy is
moving leftward.}\label{fig:flow}
\end{figure}

\begin{figure}
  \begin{center}
    \FigureFile(80mm,80mm){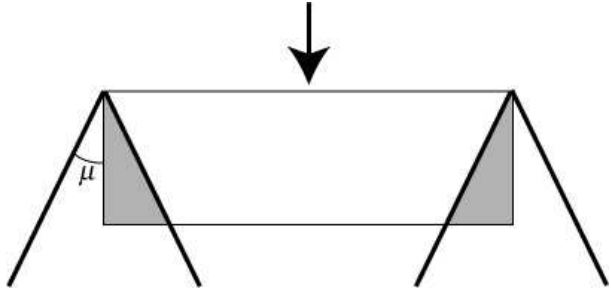}
  \end{center}
  \caption{The effect of wing ends for a rectangular wing. The influence
 of the ends is limited to the grey portions inside the Mach cones (bold
 lines)}\label{fig:foil}
\end{figure}

I consider a disk galaxy that falls into a cluster from the outside. The
galaxy should have an almost radial orbit and should be affected by
aerodynamical forces when it passes the central region of the
cluster. Suppose that a galaxy is moving in the ICM with a velocity of
$V$ (Fig.~\ref{fig:flow}). The angle of attack, $\alpha$, is defined as
the angle between the disk and the traveling direction of the galaxy. If
$\alpha$ is much smaller than $90^\circ$, ram-pressure stripping is not
strong; the ISM of the galaxy is not completely removed and makes the
surface of the disk. The ICM flows on both sides of the (remaining) ISM
disk. It is assumed that the ISM disk is thin and its radius is $r$. If
${\cal M}>1$, the lift force on the disk is given by
\begin{equation}
 F_{\rm lift}\approx 2 \rho_{\rm ICM}V^2 S \alpha/\sqrt{{\cal M}^2-1}\:,
\end{equation}
where $S=\pi r^2$ is the area of the disk \citep{lan87}. The lift force
is not sensitive to the form of the disk cross-section
\citep{lan87}. Although this equation is for an infinitely long airfoil,
the correction for a finite long one is small. The influence of the wing
ends is limited to the inside of the Mach cones with the apexes at the
wing ends and the angle of $\mu=\sin^{-1}(1/{\cal M})$
(Fig.~\ref{fig:foil}). For a disk, its portions included in the cones
are small.

For a luminous disk galaxy, the radius of the ISM disk is $r\sim 25$~kpc
if ram-pressure stripping can be ignored totally.  Considering the
ram-pressure stripping, $r\sim 5$~kpc is taken here
\citep{vol01,roe05b}. For the central region of a rich cluster similar
to the Coma cluster, the ICM temperature and density are $T_{\rm
ICM}\sim 8$~keV and $\rho_{\rm ICM}\sim 2\times 10^{-27}\rm\: g\:
cm^{-2}$, respectively. Thus, for a disk galaxy passing through the
central region with ${\cal M}\sim 2$, the lift force is
\begin{eqnarray}
 F_{\rm lift}&\approx &3.1\times 10^{34} {\rm\: dyn}
\left(\frac{\rho_{\rm ICM}}{2.3\times 10^{-27}\rm\: g\: cm^{-2}}\right)
\left(\frac{V}{3000\:\rm km\: s^{-1}}\right)^2 \nonumber\\
& &\times
\left(\frac{r}{5\rm\: kpc}\right)^2
\left(\frac{\alpha}{10^{\circ}}\right)
\left[4\left(\frac{{\cal M}}{2}\right)^2-1\right]^{-1/2}\:.
\label{eq:lift}
\end{eqnarray}
As a result, the galaxy gains extra velocity perpendicular to its
original direction of travel, $\Delta V\sim (F_{\rm lift}/M)\: t_{\rm
cross}\sim 25\rm\: km\: s^{-1}$, where $M$ ($\sim 10^{11}\: M_\odot$) is
the galaxy mass, and $t_{\rm cross}$ ($\sim 5\times 10^8$~yr) is the
time for the galaxy to cross the central region of the cluster.  Since
this is much smaller than the velocity dispersions of galaxies in rich
clusters ($\sim 1000\rm\: km\: s^{-1}$), it can be ignored. It is to be
noted that since $\alpha$ must be small enough to avoid ram-pressure
stripping \citep{roe05b} and large enough to lift the galaxy ($\sim
10^{\circ}$), the fraction of galaxies affected by the lift force
($\Delta V\sim 25\rm\: km\: s^{-1}$) is very limited. After the galaxy
reaches the apocenter far away from the cluster center, it falls back
into the cluster again. However, it is unlikely for the galaxy to have
an attack angle of $\sim 10^{\circ}$ again.

\section{Drag Force}
\label{sec:drag}

For a poor cluster with a temperature of $T_{\rm ICM}\sim 2$~keV,
ram-pressure stripping is almost ineffective for most galaxies
regardless of the attack angle $\alpha$, because the velocity of the
galaxies and thus the ram-pressure on them are small. As a result, the
radius of a ISM disk is large, and the drag force on a galaxy would be
important especially when $\alpha\sim 90^\circ$. For spherical galaxies,
the drag force has been studied, although its influence on galaxy motion
has not been discussed \citep{sha82,gae87}. For a given radius, the
drag force on a disk perpendicular to wind ($\alpha=90^\circ$) is much
larger than that on a sphere \citep{whi86}. Therefore, it is worthwhile
to revisit the problem for disk galaxies.

The drag force is given by
\begin{equation}
 F_{\rm drag}=\frac{1}{2}C_D\rho_{\rm ICM}V^2\pi r^2\:,
\end{equation}
where $C_D$ is the drag coefficient. For a disk of $\alpha=90^\circ$,
the coefficient is $C_D=1.17$, while for a sphere with the same radius,
the coefficient is $C_D\approx 0.2$--0.47 depending on the Reynolds
number \citep{whi86}. The values of $C_D$ are for incompressible fluids
(${\cal M}\lesssim 0.3$); they rise by a few tens percent at ${\cal
M}\sim 0.6$--0.8 \citep{hen76}.

I consider a disk galaxy ($M=10^{11}\: M_\odot$ and $\alpha=90^\circ$)
orbiting well-inside a poor cluster with a temperature of $T_{\rm
ICM}=2$~keV and a density of $\rho_{\rm ICM}\sim 1\times 10^{-27}\rm\:
g\: cm^{-3}$. Since the galaxy is orbiting inside the cluster and the
drag force is not as sensitive to the attack angle as the lift force,
the galaxy is constantly affected by the drag force from the
ICM. Assuming that ${\cal M}=0.8$, the drag force is
\begin{eqnarray}
 F_{\rm drag}&\approx &5.4\times 10^{33} 
{\rm\: dyn}\left(\frac{C_D}{1.2}\right)
\left(\frac{\rho_{\rm ICM}}{9\times 10^{-28}\rm\: g\:
 cm^{-3}}\right)\nonumber \\
& &\times\left(\frac{V}{580\rm\: km\: s^{-1}}\right)^2
\left(\frac{r}{10\rm\: kpc}\right)^2
\:.
\label{eq:drag}
\end{eqnarray}
Considering weak ram-pressure pressure stripping, the disk radius was
assumed to be $r=10$~kpc \citep{aba99,roe05a}. After $t=10$~Gyr, which
is the typical age of poor clusters, the galaxy velocity decreases by
$\Delta V\sim (F_{\rm drag}/M) t\sim 90\rm\: km\: s^{-1}$ and the
deviation from the orbit that the galaxy would have been in if it were
not for the ICM is $D\sim (1/2)(F_{\rm drag}/M)\: t^2\sim 0.4$~Mpc,
which is comparable to the size of poor clusters ($\sim 1$~Mpc). Note
that although ram-pressure may distort the ISM disk, it is unlikely that
the disk becomes as round as a sphere. Thus, the drag coefficient,
$C_D$, is at least larger than 0.2--0.47.

\section{Discussion}

I have shown that while the lift force on disk galaxies can be ignored,
the drag force could not be ignored for face-on disk galaxies in poor
clusters.

Equations~(\ref{eq:lift}) and~(\ref{eq:drag}) show that the strengths of
the lift and drag forces are sensitive to the radius of the ISM disk,
$r$. In the above estimates, gas supply from stars in the galaxy was not
considered. If the gas ejected from the stars is added to the ISM, $r$
should become larger \citep{gae87}. It is likely that active disk
galaxies, in which large amount of gas is being released from the stars,
have much larger $r$ and are more affected by the aerodynamic forces. It
is also to be noted that if the velocity or pressure distribution of the
ICM around a disk galaxy is measured, the aerodynamic forces can be
constrained.  The observations of stripped ISM will be useful to obtain
the velocity field around a disk galaxy \citep{yos04,che06}. X-ray
observations will tell us the pressure distribution around a galaxy
\citep{vik01}.

The models presented here are very simple.  In an actual galaxy, the
situation would be rather delicate.  One example is how the lift or drag
would operate on the collisionless stars and dark matter particles, as
required to alter the orbit of the whole galaxy. The connecting force is
gravity. If the gas disk is lifted or dragged off the stellar disk, and
the center of the dark matter halo, then both will pull back on the gas
disk. This would communicate the lift or drag force to all
components. Too much separation, and the restoring gravity force will be
weakened and the gas will be stripped, too little and nothing will
happen. In order to study the process quantitatively, numerical
simulations would be required.

\vspace{5mm}

I am grateful to C. Struck for useful comments.  Y.~F.\ was supported in
part by a Grant-in-Aid from the Ministry of Education, Culture, Sports,
Science, and Technology of Japan (17740182).

\end{document}